\documentclass{article}
\usepackage{jcappub}
\usepackage{graphicx}
\usepackage{amsmath}

\newcommand{\be}{\begin{equation}}
\newcommand{\ee}{\end{equation}}
\newcommand{\bea}{\begin{eqnarray}}
\newcommand{\eea}{\end{eqnarray}}

\def \omms   {\Omega_m}
\def \omm  {\Omega_{0 {\rm m}}}

\renewcommand{\(}{\left(}

\newcommand{\lcdm }{$\Lambda$CDM }
\title{Dynamical dark energy models with singularities in the view of the forthcoming  results of the growth observations}

\author[a]{Tomasz Denkiewicz}
\emailAdd{atomekd@wmf.univ.szczecin.pl}
\affiliation[a]{\it Institute of Physics, University of Szczecin, Wielkopolska 15,
          70-451 Szczecin, Poland}
\affiliation[a]{\it Copernicus Center for Interdisciplinary Studies,
S{\l }awkowska 17, 31-016 Krak\'ow, Poland}
\date{\today}
\input epsf
\abstract{The question of the origin of the recent acceleration of the Universes expansion is still pending. What is making the situation even worst, it is impossible to distinguish the vast majority of the proposed models of the dynamical dark energy and modified gravity from the $\Lambda CDM$ in view of recent geometrical and dynamical, observational data. On the other hand on scales much smaller than the present Hubble scale, there are differences in the growth of the matter perturbations for different modes of the perturbations in the $\Lambda CDM$. In the view of the new planned observations that will give insight into the perturbations of the dark sector this issue is being worth of further investigation. We analyze the evolution of the dark matter perturbations in the dynamical dark energy models with the singularities, such as the sudden future singularity and the finite scale factor singularity. We employ the Newtonian gauge formulation for derivation of the perturbation equations for the growth function. We abandon the sub-Hubble approximation, what leads to the scale dependent solutions for the perturbations. Treating the growth function as a scale dependent allows to differentiate the dynamical dark energy models with the singularities and the dynamical dark energy models and the $\Lambda CDM$. The new data constraining growth of the perturbations will be able to rule out the whole range of the values of the parameters allowed by the present data of the dynamical dark energy models with the sudden future singularity and the finite scale factor singularity.
}

\begin{document}
\maketitle

\section{Introduction}
\setcounter{equation}{0}
Current observations, like the measurements of the luminosity distance with the supernovae type Ia (SnIa) \cite{1998AJ,1999ApJ}, the scaled distance to the last scattering surface ($\mathcal{R}$) \cite{9702100v2} or the baryon acoustic oscillations (BAO) \cite{Eisenstein} are all geometrical probes, which are unable to distinguish between the models of the accelerated expansion and the modified gravity. Although those tests are probes of the dark energy and give the information about $H(z)$, but those do not give much insight into the properties of the dark energy. Often, those are also not necessary the tests, which are sufficient to discriminate between the dynamical dark energy models themselves and between them and the $\Lambda CDM$. 
Other type of the tests are the dynamical probes, which allow to probe the growth of the matter perturbations. The data up to now do not provide very restrictive constraints onto the parameters of the models, which are currently on the market, but to some extent the combination of the data for $H(z)$ and the density contrast $\delta(z)=\delta \rho(z)/\rho(z)$ allow to differentiate between the dark energy models and the modified gravity such as $f(R)$ models  and Gvali-Gabadadze-Porrati (DGP) model and the $\Lambda CDM$ \cite{0908.2669v1,0809.3374v2,astro-ph/0701317v2,arXiv:1207.1009}.\\
Lack of a firm explanation of the origin of the present Universes accelerated expansion led to the formulation of the different scenarios of the modified gravity models and the models within the general relativity, which foreseen different types of the singularities in a finite time, in the course of the Universes evolution. A big-rip (BR), a sudden future singularity (SFS) \cite{barrow04, barrow042, gr-qc/0410033v3,hep-th/0505215v4}, a generalized sudden future singularity (GSFS), a finite scale factor singularity (FSF) \cite{0910.0023v1,hep-th/0505215v4} or a big-separation singularity (BS) \cite{not2005,hep-th/0505215v4} and $w$-singularities \cite{0902.3107v3} gained some attention among the cosmologists community.\\
We are dealing with an issue of the observational testing, constraining, differentiating and hopefully falsifying some of those scenarios. The forthcoming data from the planned observations, which intend to give the constrains onto the growth function, are very promising \cite{astro-ph/0703191v2}. Future galaxy week lensing experiments like the Dark Energy Survey (DES) \cite{des} and Euclid \cite{euclid} and such an observation like SKA project \cite{SKA} potentially will be able to discriminate between the cosmological constant $\Lambda$CDM and evolving dark energy scenarios. The forthcoming observations of the cosmic background radiation in the microwave to far-infrared bands in the polarization and the amplitude, as the Polarized Radiation and Imaging Spectroscopy (PRISM) \cite{prism} and the very high precision measurements of the polarization of the microwave sky by the Cosmic Origins Explorer (CoRE) satelite \cite{core} will improve the constraints onto the dark sector. \\
In this work we show, that with the new data constraining the growth function, it will be possible to discriminate between the different models of the dynamical dark energy, which foreseen different kinds of the singularities in the course of the evolution of the universe to appear. Also it will be possible to discriminate between those models and the $\Lambda CDM$. For that, it is necessary to go beyond the scale independent approximation for the equation for the matter density evolution (eq.~\ref{dcesi}). It is needed to consider the full set of the scale dependent equations (\ref{gr1}-\ref{gr3}) or, in the case, in which one is restricted to the sub-Hubble scales, the growth rate parametrization, which is scale dependent (eq.~\ref{approxfk}). Evolution of the linear density perturbations for FSFS, Big Rip, SFS, FSFS and Pseudo-Rip was a subject of some of the previous works \cite{1202.3280v2,AO,1411.6169v2}. \\
In the following Section \ref{perturbations} we introduce the background equations, and the perturbed Einstein equations in a Newtonian frame \cite{astro-ph/9506072v1}. In the Section \ref{themodels}, we give a short description of the dynamical dark energy models with the SFS and the FSFS. In the last Section \ref{lastS}, we give the results for the dark matter perturbation evolution, the growth function, in models with the singularities, for the different scales of the perturbation modes and as the reference model we use the $\Lambda CDM$. We follow up with the conclusions.

\section{Scale dependent growth function}\label{perturbations}
\setcounter{equation}{0}
The growth function is a useful tool as a probe of the dynamics of the Universes expansion. It is often used with the scale independent approximation for the perturbation modes. It is parametrized, in the following form, with the definition of the growth rate $f$ given by the following expression:
\be
f(a)\equiv \frac{d\ln \delta_{m}}{d\ln a}=\omms(a)^\gamma, \label{fofa} 
\ee
where $a$ is the scale factor and with the introduction of the growth index $\gamma$, with
\be
\omms (a) \equiv \frac{H_0^2 \omm a^{-3}}{H(a)^2} \label{omadef}.
\ee
This parametrization is an approximation to the scale independent solution of the equation for the growth rate:
\be f' + f^2 + f(\frac{\dot H}{H^2}+2)=\frac{3}{2} \omms \label{greqlna}, \ee  where $' =d/dlna$, which is obtained from the density contrast evolution equation:
 \be {\ddot \delta}_m + 2 H {\dot \delta}_m - 4\pi G \rho_m \delta_m =0, \label{dcesi} \ee
 with the change of the variable  from $t$ to $\ln a$, where  an overdot denotes the derivative with respect to the time and $\rho_m$ is the matter density. It was found that $\gamma=6/11$ for the $\Lambda CDM$ \cite{astro-ph/9804015v1,astro-ph/0507263v2,astro-ph/0701317v2}. For the dark energy models with the slowly varying equation of state the solution is approximated with the equation (\ref{fofa}), where
\be
\gamma=\frac{3(w_0-1)}{6w_0-5},
\ee
with $w_0\equiv w(a)=p/\rho$, which for the $\Lambda CDM$ is, $w_0=-1$. It has been shown, that for the slowly varying dark energy models the $\gamma$ index is not strongly time dependent. If it varies at all, it does at a few percent level \cite{1002.3030v1,0903.5296v2}. It was found, that for the modified gravity models, it was not longer the case. The $\gamma$ may vary and for the well known case of the DGP model,  $\gamma\simeq 0.68$ and a better approximation to the solution of the full set of equations (\ref{gr1}-\ref{gr3}) is obtained when the redshift dependence is taken into account. The different parametrizations for the growth index were proposed \cite{0905.2470v2,astro-ph/0701317v2,0802.4196v4,0808.1316v2,arXiv:1207.1009}. \\
Having in mind the forthcoming data for the growth it seems, that one should be careful using also the scale independent approximation during the derivation of eq. (\ref{dcesi}). It was found \cite{0808.2689v4,0903.5296v2,1002.3030v1}, that the exact solution for the full set of the scale dependent equations, for the growth, shows a scale dependence on the scales larger than $100h^{-1}Mpc$ for the $\Lambda CDM$. It was argued, that the scale invariant approximation brakes down because of the sub-Hubble scale assumption for the perturbation modes is used and this one breaks down already for the scales around $200h^{-1}$ in the early stages. No matter that recently the Hubble scale is around $3000h^{-1}Mpc$ . This is the case for the $\Lambda CDM$ and the singularity scenarios which are considered in this work \cite{0808.2689v4,0903.5296v2,1002.3030v1,1411.6169v2}. \\
It was already shown \cite{1411.6169v2}, that the models of the dynamical dark energy, with the singularities, differ with respect to the mode wavenumber, for which the amplitude of the dark energy perturbation is of the order of the dark matter perturbation amplitude. It is important to have in mind, that for some wavelengths, this is model dependent, the perturbations in the dark energy couples with the dark matter perturbations and can not be ignored.\\
In this work we consider the perturbations in the Newtonian gauge. In order to test the scale dependent approximation in the singular models in the view of the new forthcoming data from DES and Euclid, 
we defy the assumption of the sub-Hubble scale of the perturbation modes. We consider the scale dependent solution to the full set of the perturbation equations for the FSFS and SFS models and for the comparison we use the $\Lambda CDM$ as a reference model.\\
The perturbed metric in the Newtonian gauge with the assumption of the lack of the anisotropic stress takes the form:
\begin{equation}
 ds^2 = -(1+2\Phi) dt^2 + (1-2\Phi)a^2\gamma_{ij}dx^i dx^j,
\end{equation}
where $\gamma_{ij}$ is the spatial part of the metric and  $\Phi$ is the Newtonian potential. With the assumption, that the universe is flat and filled only with the pressureless, nonrelativistic dark matter, $\rho_m$, and an exotic fluid (which we call dark energy with the energy density $\rho_{de}$), the evolution is governed by the background Firedmann equations: 
\bea H^2 &=& \frac{8\pi G}{3}(\rho_m +\rho_{de}), \label{fried} \\ {\dot \rho} &=& -3H (\rho + p) \label{cont} \eea    
and the perturbed up to linear order Einstein equations in the Newtonian gauge, that result with the set of equations as follows:
\bea
\label{gr1}\ddot{\Phi}&=&-4H\dot{\Phi}+8\pi G \rho_{de} w_{de}\Phi,\\
\label{gr2}\dot{\delta}&=&3\dot{\Phi}+\frac{k^2}{a^2}v_{f},\\
\label{gr3}\dot{v}_{f}&=&-\Phi,
\eea
with the following constraint equations:
\bea
\label{constraint1}3H(H\Phi+\dot{\Phi})+\frac{k^2}{a^2}\Phi&=&-4\pi G\delta\rho_m,\\
\label{constraint2}(H\Phi+\dot{\Phi})&=&-4\pi G\rho_m v_{f}.
\eea
Here $v_f=-va$, and $v$ is the velocity potential for the dark matter, $k$ is the wavenumber.
With the sub-Hubble approximation $k^2/a^2>>H^2$ and a slowly varying gravitational potential one arrives with the scale independent equation for the matter density contrast evolution in the form (\ref{dcesi}). When the sub-Hubble approximation is relaxed and the approximation of a slowly varying Newtonian potential is hold one gets the scale dependent evolution for the $\delta_m$ in the following form \cite{1002.3030v1,0903.5296v2}:
\bea {\ddot \delta_m} + 2 H {\dot \delta_m} - \frac{4\pi G \rho_m \delta_m}{1+\xi(a,k)}=0,\label{delsd} \eea
where \bea \xi(a,k)=\frac{3 a^2 H(a)^2}{k^2}. \label{xidef} \eea
The equation (\ref{delsd}) may be expressed in terms of the growth factor, $f=\frac{d\ln \delta_m}{d\ln a}$ in the form
\bea f' + f^2 + \(2-\frac{3}{2} \omms(a)\)f=\frac{3}{2}\frac{\omms(a)}{1+\xi(a,k)},\label{fsd}  \eea
where $'\equiv \frac{d}{d\ln a}$ and we have assumed \lcdm for $H(a)$.
The approximate solution to the equation (\ref{fsd}) can be parametrized as following:
\be 
f(k,a)=\frac{\omms(a)^\gamma}{1+\frac{3H_0^2\omm}{ak^2}}.\label{approxfk}
\ee
For the sub-Hubble scales $\xi(k,a)\rightarrow 0$ and the equation (\ref{fsd}) reduces to the (\ref{greqlna}) whose solution is well approximated by the (\ref{fofa}) with the $\gamma=\frac{6}{11}$. \\
It was shown, that the equation (\ref{delsd}) provides a better approximation to the full general relativistic system (\ref{gr1}-\ref{gr3}) up to the horizon scales \cite{1002.3030v1,0903.5296v2} and for a larger scales one can not ignore the change in the time of the potential, $\Phi$.
\section{FSFS and SFS as the dynamical dark energy candidates}\label{themodels}
\setcounter{equation}{0}

The SFS and FSFS show up within the framework of the Einstein-Friedmann cosmology governed by the
standard field equations
(\ref{fried})
and the energy-momentum conservation law (\ref{cont}).
%

We get the SFS and FSFS scenarios with the scale factor in the following form:
\be
\label{sf2} a(t) = a_s \left[b + \left(1 - b \right) \left( \frac{t}{t_s} \right)^m - b \left( 1 - \frac{t}{t_s} \right)^n \right],
\ee
the appropriate choice of the constants is needed $b, t_s, a_s, m,n$ \cite{barrow04,DHD}. For both cases the SFS as well as the FSFS model the evolution starts with the standard big-bang singularity at $t=0$, for $a=0$, and evolves to an exotic singularity for $t=t_s$, where $a=a_s\equiv a(t_s)$ is a constant. \\
Accelerated expansion in an SFS universe, is assured with the negative $b$, ($b<0$), and as for the FSFS universe, $b$ has to be positive ($b>0$). In order to have the SFS $n$ parameter has to be within the range $1<n<2$, while for an FSFS, $n$ has to obey $0<n<1$. For the SFS at $t=t_s$, $a \to a_s$, $\varrho \to \varrho_s=$ const., $p \to \infty$, while for an FSFS the energy density $\rho$ also diverges and one has: for $t\rightarrow t_s$, $a\rightarrow a_s$, $\rho\rightarrow\infty$, and $p \rightarrow \infty$, where $a_s,\ t_s$, are constants and $a_s\neq 0$.\\
%
In both scenarios the nonrelativistic matter scales as $a^{-3}$, i.e.
\be
\rho_m=\Omega_{m0}\rho_0\left(\frac{a_0}{a}\right)^3,
\ee
and the evolution of the exotic (dark energy) fluid $\rho_{de}$, can be determined by taking the difference between the total energy density $\rho$, and the energy density of the nonrelativistic matter, i.e.
\be
\rho_{de}=\rho-\rho_m~~.
\ee
In those scenarios $\rho_{de}$ component of the content of the Universe is responsible for an exotic singularity at $t\rightarrow t_s$. The dimensionless energy densities are defined in a standard way as
\be
\Omega_m=\frac{\rho_m}{\rho}, \hspace{0.3cm} \Omega_{de}=\frac{\rho_{de}}{\rho}.
\ee
For a dimensionless exotic dark energy density we have the following expression
\be
\label{Omde}
\Omega_{de}=1-\Omega_{m0}\frac{H_0^2}{H^2(t)}\left(\frac{a_0}{a(t)}\right)^3=1-\Omega_{m}.
\ee
The barotropic index of the equation of state for the dark energy is defined as
\be
\label{wde}
w_{de}=p_{de}/ \rho_{de}.
\ee
Singularity scenarios considered in this work were also tested as the candidates for the dynamical fine structure constant cosmologies \cite{alpha}. In that approach the dark energy is sourced with a scalar field which couples with the electromagnetic sector of the theory. The knowledge about the effective evolution of the dark energy density and the dark energy equation of state evolution is sufficient to estimate the resulting fine structure constant evolution.
\section{Results and conclusions}\label{lastS}\setcounter{equation}{0}
\begin{figure}
\begin{center}
  \resizebox{93mm}{!}{\includegraphics{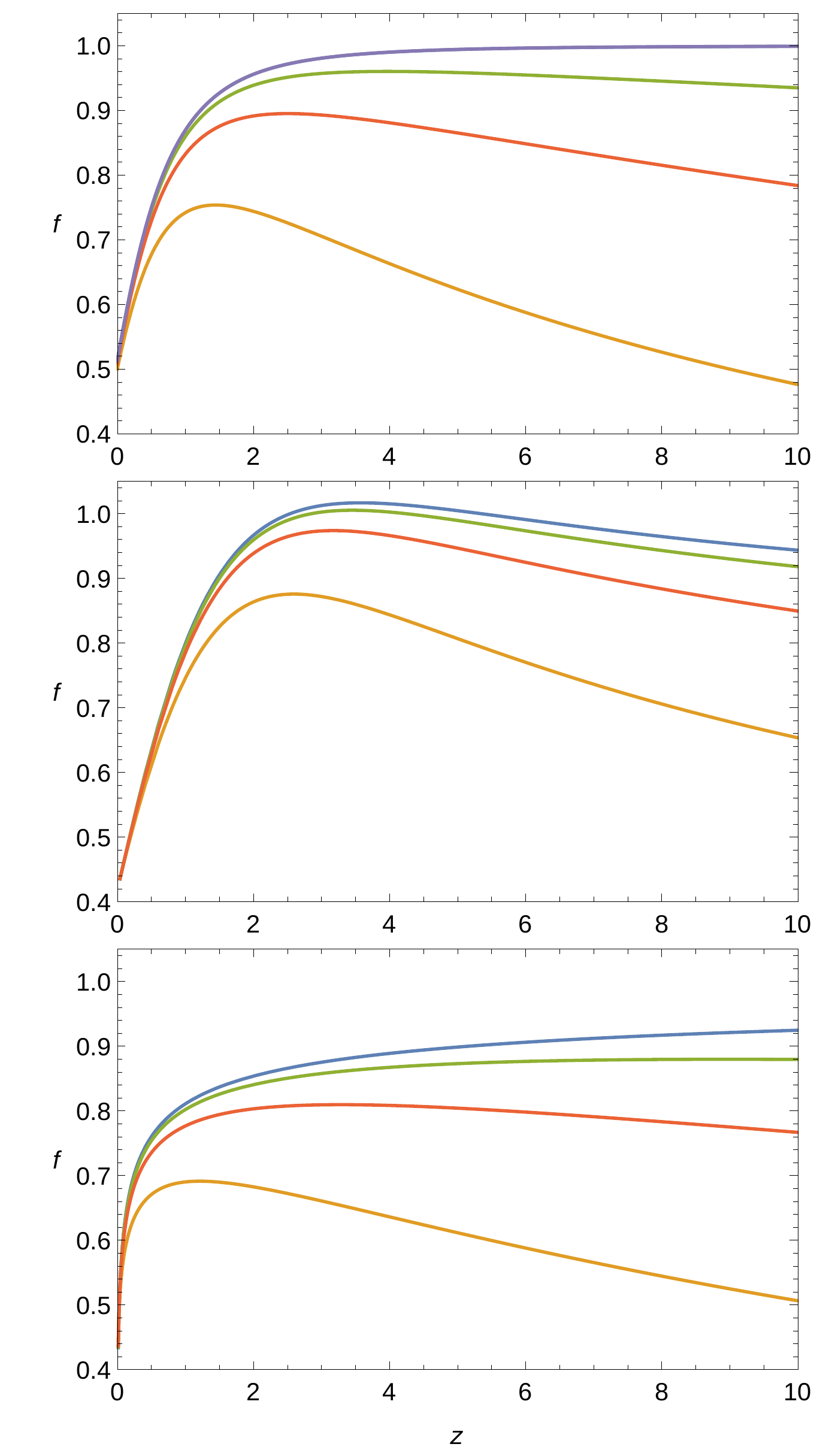}}
    \caption{The plot of the growth factor, $f$  for $\Lambda CDM$, the SFS and the FSFS scenarios from the top to the bottom respectively. In each plot, the growth factor for four different modes of perturbation is plotted, it is $k=0.1hMpc^{-1}$, $k=0.004hMpc^{-1}$, $k=0.002hMpc^{-1}$, $k=0.001hMpc^{-1}$ from the top to the bottom respectively. The values of the parameters for the models with the future singularity are taken as for the SFS model: $m=0.749$, $n=1.99$, $b=-0.45$, $y_0=0.77$, for the FSFS: $m=2/3$, $n=0.7$, $b=0.24$, $y_0=0.96$.}\label{figura}
\end{center}
\end{figure}
Being interested in the detailed information on the evolution of the perturbations in the dark sector, we are interested in the solution to the full set of the eqs. (\ref{gr1}-\ref{gr2}). For the SFS and the FSFS scenarios, with the model parameters given in the caption of the Fig. \ref{figura}, we plot the growth function as a function of the redshift. The models that were chosen fulfil constraints given by the geometrical probes like the SNIa, BAO, shift parameter, Hubble parameter evolution \cite{DHD,FSF,DDGH,rd,GHDD}. For the different wavelengths of the perturbation modes (given in the caption of the figure) we plot the growth function for the singular scenarios and the $\Lambda CDM$ as a reference model. The values of the growth factor for a given wavelength differ from the model to the model. In addition it can be seen that the discrepancies between the values of the growth function for different wavelengths of perturbations differ from the model to the model. From the Fig. \ref{figura}. it is clearly visible that for all of the investigated models for the mode, $k\simeq0.002hMpc^{-1}$ the discrepancy between it and the shorter wavelength mode, for redshift, $z\simeq5$, is around $10\%$ and it grows with the redshift. In general having a look at greater redshifts, in this domain, will put more stringent constraints onto the dynamical dark energy models. The results for the growth function are model parameter dependent and the different dynamical dark energy models of this type give disparate values.\\
For the quintessence models the perturbations in the dark energy play a role only on scales comparable to Hubble scale. But for models for which the scalar field is not canonical or for dynamical dark energy models in which dark energy speed of sound is $c_s<<1$ the perturbations in dark energy can grow as perturbations in the dark matter for smaller scale \cite{0806.3461v1,0909.0007v2,1303.0414v2,1411.6169v2}.\\ 
As a result, in a view of the new forthcoming data from the observations of the polarization of the CMB like CoRE and PRISM, it is interesting to explore the details of the evolution of the perturbations in the dark sector. Additionally the data from weak lensing experiments like Euclid and DES will put much stringent constraints onto the dynamical dark energy models \cite{astro-ph/0703191v2}. Having data for the growth of the perturbations for the bigger wavelengths and for the bigger redshifts will put more constrains onto the class of the dynamical dark energy models with the singularities like the sudden future singularity and the finite scale factor singularity scenarios.


\newpage
\section{Acknowledgements}
\indent We are grateful to Leandros Perivolaropoulos for his kind advice concerning the work. This project was financed by the Polish National Science Center Grant DEC-2012/06/A/ST2/00395.
\appendix

\bibliography{growthscale3}{}
\bibliographystyle{JHEP}

\end{document}